\documentclass[aps,preprint]{revtex4}
\usepackage[dvips]{graphics,graphicx}
\begin{document}

\title{Simulating cyclotron-Bloch dynamics of a charged particle in a 2D lattice by means of cold atoms in driven quasi 1D optical lattices}
\author{Andrey R. Kolovsky$^{1,2}$}
\affiliation{$^1$ Kirensky Institute of Physics, 660036 Krasnoyarsk, Russia}
\affiliation{$^2$ Siberian Federal University, 660036 Krasnoyarsk, Russia}
\date{\today}

\begin{abstract}
Quantum dynamics of a charged particle in a 2D lattice subject to magnetic and electric fields is a rather complicated interplay between cyclotron oscillations (the case of vanishing electric field) and Bloch oscillations (zero magnetic field), details of which has not yet been completely understood. In the present work we suggest to study this problem by using cold atoms in optical lattices. We introduce a 1D model which can be easily realized in laboratory experiments with quasi 1D optical lattices and show that this model captures many features of the cyclotron-Bloch dynamics of the quantum particle in 2D square lattices.
\end{abstract}
\maketitle

\section{Introduction}

A charged particle in 2D periodic potentials, which is subjected to in-plane electric field and normal to the plane magnetic field is a problem of lasting fundamental interest because of its relation to the quantum Hall effect. This problem was considered in several different approximations in the solid-state physics literature with the main emphasis on the energy spectrum of the system or, more precisely, on the electron density of states, which is a measurable quantity. Among these approximations the most popular is the tight-binding approximation, which amounts to a truncation of the Hilbert space of the single-particle Hamiltonian to the lowest Bloch band. In the case of zero electric field this approximation results in the celebrated Hofstadter's butterfly spectrum \cite{Hofs76}, which is parametrized by the Peierls phase $\alpha$ -- number of the magnetic flux quanta through the unit cell of the lattice \cite{Peie33}. In the opposite case of zero magnetic field the spectrum is the so-called Wannier-Stark ladder \cite{Naka93,51}, which is parametrized by the angle $\theta$ between the electric field vector and the crystallographic axis of the lattice. The case when both fields are present is more subtle and, to the best of our knowledge, was analyzed for the first time only in 1995 \cite{Naka95}.

Complementary to the spectral problem is the wave-packet dynamics of the particle. The main question one addresses here is whether a localized packet remains localized in the course of time or it spreads over the lattice. If for electrons in a solid crystal this problem is of pure academic interest, it appears to be of experimental relevence for cold atoms in optical lattices because in this system the wave-packet dynamics can be easily tracked by taking a picture of the atomic cloud after a given evolution time. A recent example is Ref.~\cite{Roat08} where the authors realized the Aubry-Andre model \cite{Aubr80} (which coincides with Harper's Hamiltonian \cite{Harp55} for the model parameter $\lambda=2$)  by loading cold atoms into the quasiperiodic 1D optical lattice. It was confirmed that the atomic cloud remains localized for large $\lambda$ and spreads over the lattice for small $\lambda$.  

Present research in cold atoms physics is also focused on the problem of generating synthetic magnetic fields, which could impart a Lorenz-like force to otherwise neutral atoms in motion (see, for example, the recent paper \cite{84} and references therein). Since the electric field for cold atoms is easily mimicked by, for example, the gravitational force, this will open a perspective for studying the 2D wave-packet dynamics in the Hall configuration. Theoretically,  this problem was analyzed in much details in our recent publications \cite{85,preprint}. The main message of the present work is that many (although not all) theoretical predictions of Refs.~\cite{85,preprint} can be verified  by using the driven 1D lattice instead of the 2D lattice with an artificial magnetic field. The proposed experiment is a modification of the laboratory setup \cite{Roat08}, where one substitutes one of the stationary lattices by a moving lattice.  

\section{Wave-packet dynamics in the Hall configuration}

To make the paper self-consistent we summarize the results of Refs.~\cite{85,preprint}. Using the tight-binding approximation the 2D wave-packet dynamics in the Hall configuration is governed by the following Schr\"odinger equation:
\begin{equation}
\label{1}
i\hbar\dot{\psi}_{l,m}=
-\frac{J_x}{2}\left(\psi_{l+1,m}  +  h.c. \right)
-\frac{J_y}{2}\left(e^{i2\pi\alpha l} \psi_{l,m+1} + h.c. \right) 
+ea(F_x l+F_y m) \psi_{l,m} \;.
\end{equation}
In this equation $\psi_{l,m}$ are the wave function probability amplitudes for the lattice site $(l,m)$, $J_x$ and $J_y$ the hopping matrix elements along the $x$ and $y$ axis (in what follows we assume $J_x=J_y\equiv J$ for simplicity),  $e$ the charge, $a$ the lattice period, $F_{x}$ and $F_y$ components of the electric field vector ${\bf F}$, $\alpha=eBa^2/hc$ the Peierls phase, and we use the Landau gauge ${\bf A}=B(0,x,0)$ for the vector potential.

The main conclusion of Ref.~\cite{85} is that the system (\ref{1}) has two qualitatively different dynamical regimes, depending on the inequality relation between the electric field magnitude and the quantity $F_{cr}=2\pi\alpha J/ea$.
%
In the strong field regime, $F>F_{cr}$, the time evolution of a localized wave-packet is defined by the Bloch dynamics. For vanishing magnetic field these would be Bloch oscillations, where the packet oscillates near its initial position with the Bloch frequencies $\omega_{x,y}=eaF_{x,y}/\hbar$ and amplitudes proportional to $J_{x,y}/eaF_{x,y}$. The obvious exception from this oscillatory behavior  is the case where the vector ${\bf F}$ points the $x$ or $y$ direction. Here the packet spreads ballistically in the orthogonal to the field direction, with the rate defined by the hopping matrix element. A finite magnetic field (nonzero $\alpha$) `generalizes' this exclusion to the cases where the vector ${\bf F}$ points rational directions, i.e., $F_x/F_y=r/q$ with $r,q$ being co-prime numbers \cite{preprint}. However, now the rate of ballistic spreading in the orthogonal direction is suppressed by the factor proportional to $(J/eaF)^{r+q-1}$. In practice this functional dependence  of the suppression coefficient implies that the wave packet spreading can be detected only for rational directions with small prime numbers $r$ and $q$.

In the opposite limit of weak electric fields, $F<F_{cr}$, the time evolution of a localized wave-packet is defined by the cyclotron dynamics. Namely, the packet moves in the orthogonal to ${\bf F}$ direction with the drift velocity $v^*$,
\begin{equation}
\label{3}
v^*=ea^2F/h\alpha=Fc/B \;,
\end{equation}
in close analogy with a charged particle in free space under the effect of the crossing electric and magnetic fields. However, the presence of the lattice imposes a restriction: this regime occurs only for the subspace of initial conditions spanned by the family of transporting states. For generic initial conditions the packet typically splits into several sub-packets moving in the orthogonal direction with different (both positive and negative) velocities.

\section{The 1D approximation}

In this section we introduce the 1D approximation to the above 2D problem.
First we use the unitary transformation, $\psi_{l,m}(t)\rightarrow \exp[-i(\omega_x l +\omega_y m)t]\psi_{l,m}(t)$, after which the electric field appears as periodic  driving of the system with Bloch frequencies $\omega_x$ and $\omega_y$:
\begin{equation}
\label{4}
i\hbar\dot{\psi}_{l,m}=
-\frac{J_x}{2}\left(e^{-i\omega_x t} \psi_{l+1,m}  +  h.c. \right)
-\frac{J_y}{2}\left(e^{i(2\pi\alpha l-\omega_y t)}\psi_{l,m+1} + h.c. \right)  \;.
\end{equation}
Let us now assume a situation where the wave function is uniform along the $y$ axis, i.e., $\psi_{l,m}(t)=L^{-1/2}b_l(t)$. This reduces  (\ref{4}) to the following 1D Schr\"odinger equation for the complex amplitude $b_l$:
\begin{equation}
\label{5}
i\hbar\dot{b}_l=-\frac{J_x}{2}(e^{-i\omega_x t} b_{l+1}+h.c.) - J_y\cos(2\pi\alpha l-\omega_yt)b_l  \;.
\end{equation}
Although the above assumption is rather specific, it was shown in Ref.~\cite{85} that $|b_l(t)|^2$ well approximates the integrated probability $P_l(t)=\sum_m |\psi_{l,m}(t)|^2$ also in the case of a localized 2D wave packet, if its size exceeds the magnetic period $d=a/\alpha$. Thus we can use the results of Refs.~\cite{85,preprint} to explain dynamical regimes of the system (\ref{5}) and, vice versa, to verify theoretical predictions of the cited papers by studying the wave-packet dynamics of this one-dimensional system. Note that in terms of Eq.~(\ref{5}) the weak and strong field regimes correspond to slow driving, $\omega<\omega_{cr}$, 
\begin{equation}
\label{6}
\omega=\sqrt{\omega_x^2+\omega_y^2}  \;,\quad  \omega_{cr}=2\pi\alpha J/\hbar \;,
\end{equation}
and fast driving, $\omega>\omega_{cr}$, respectively.

The system (\ref{5}) coincides with that studied in the laboratory experiment \cite{Roat08} with  two minor modifications. First, now the secondary (in terminology of  Ref.~\cite{Roat08}) lattice moves with the constant velocity relative to the primary lattice. This can be done by linearly chirping the frequencies of two counter-propagating waves which form the secondary lattice \cite{remark1}. Second, the hopping term in Eq.~(\ref{5}) contains an oscillatory phase. This phase can be introduced by the same techniques which one uses to induce Bloch oscillations of cold atoms (for example, by employing the gravitational force for vertically oriented lattices).

\section{The regime of slow driving}

\begin{figure}[b!]
\center
\includegraphics[width=10cm,clip]{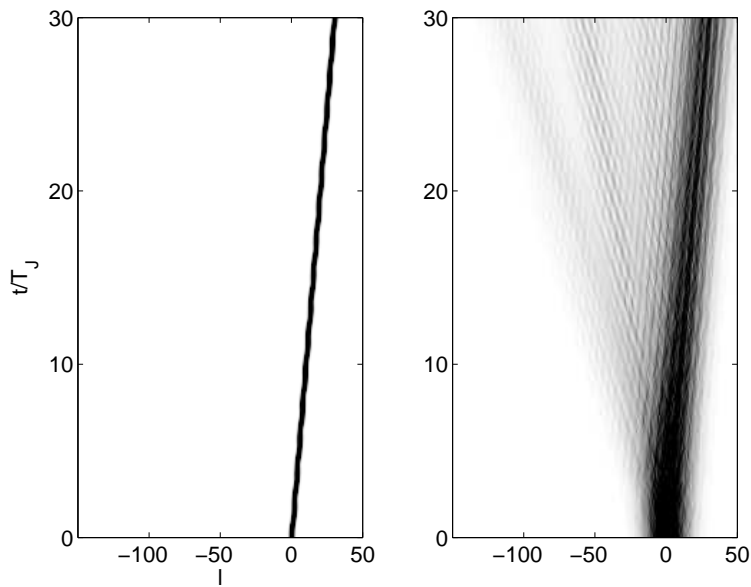}
\caption{Space-time plot of the dynamics of the narrow coherent (left panel) and a wide incoherent (right panel) Gaussian wave-packets.  Parameter are $J_x=J_y=1$, $\alpha=1/10$, $\omega=0.1$, and $\omega_x/\omega_y=0$. The time is measured in units of the tunneling period $T_J=h/J$.}
\label{fig1}
\end{figure}

A way to check that the 1D system (\ref{5}) correctly captures the features of the 2D system (\ref{1}) in the weak field regime (which now assumes  $\omega<\omega_{cr}$) is to propagate the transporting state. In the 2D lattice it moves with the drift velocity (\ref{3}) in the orthogonal to ${\bf F}$ direction. For the considered system (\ref{5}) this means that one can construct a coherent wave packet, which will travel across  the lattice with the constant velocity 
\begin{equation}
\label{7}
v=a\omega_y/2\pi\alpha \;.
\end{equation}
Figure \ref{fig1}(a) shows the result of numerical simulation for the initial Gaussian wave packet ,
\begin{displaymath}
b_l(t=0) \sim \exp(-l^2/2\sigma^2) \;,
\end{displaymath}
with the width $\sigma=(2\pi\alpha\sqrt{J_y/J_x})^{-1/2}$. This packet approximates the ground Wannier state for the potential $V(l)=-J_y\cos(2\pi\alpha l)$ and is a 1D analogue of the 2D transporting state with the same dispersion in the orthogonal direction \cite{remark2}. It is seen in Fig.~\ref{1}(a) that the packet indeed moves with the velocity given in Eq.(\ref{7}).

To create the narrow coherent atomic wave-packet might be a problem in a laboratory experiment. For this reason from now on we focus on the case of thermal atomic cloud, which corresponds to a wide incoherent Gaussian wave packet. We simulate the dynamics of this incoherent packet by assigning random phases to probability amplitudes of the initial Gaussian packet and averaging the result over different realizations of these random phases. The right panel in Fig.~\ref{fig1} shows the typical evolution of a wide incoherent packet. Unlike the case of a narrow coherent packet, now the first moment $M_1=\sum l |b_l|^2$ remains constant while the dispersion $\sigma(t)=\sqrt{M_2-M_1^2}$  grows linearly in the limit of large times.

\section{The regime of fast driving}

\begin{figure}
\center
\includegraphics[width=10cm,clip]{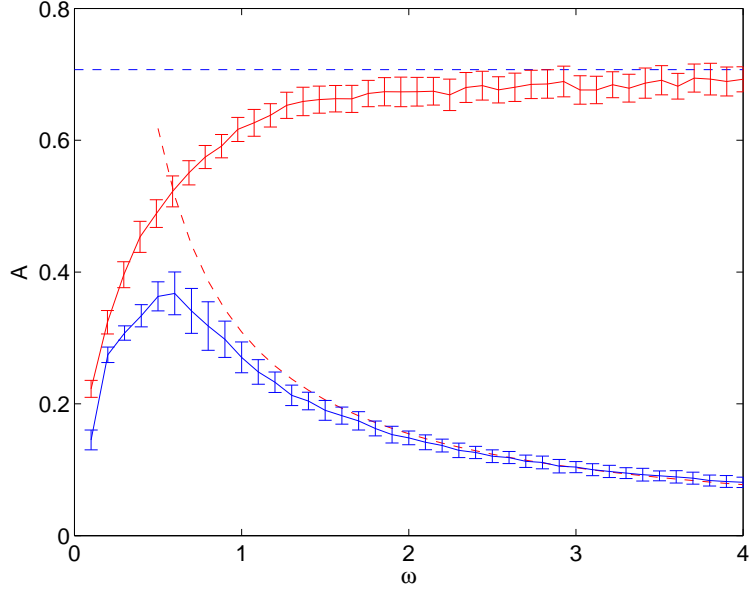}
\caption{The proportionality coefficient $A=A(\omega)$ for the asymptotic linear growth of the wave-packet dispersion, $\sigma(t)\approx At/T_J$, for $\omega_x/\omega_y=0$ and $\omega_x/\omega_y=1$. The dashed lines are analytical estimates for large $\omega$ according to Ref.~\cite{preprint}.}
\label{fig2}
\end{figure}

\begin{figure}
\center
\includegraphics[width=10cm,clip]{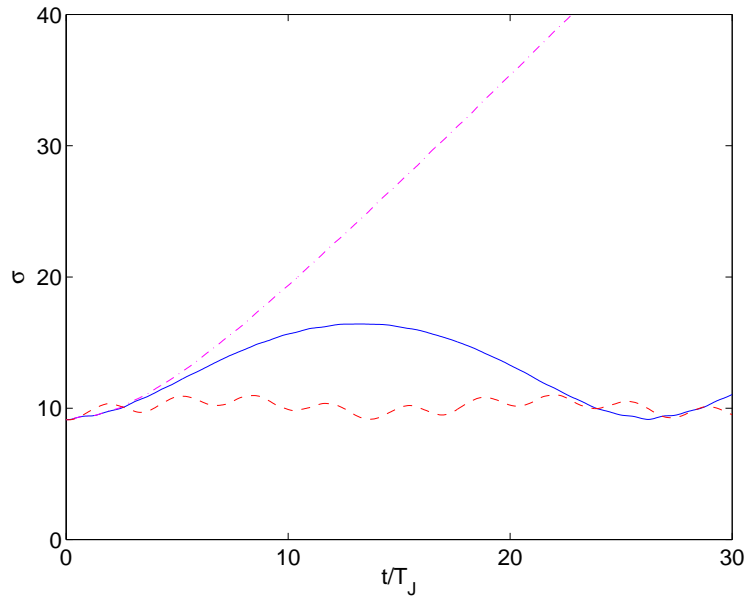}
\caption{The wave packet dispersion as the function of time for $\omega_x/\omega_y=1$ (dash-dotted line), $\omega_x/\omega_y=18/19$ (solid line), and $\omega_x/\omega_y=(\sqrt{5}-1)/4$ (dashed line). The other parameters are the same as in Fig.~1(b) yet $\omega=1$.}
\label{fig3}
\end{figure}

The characteristic feature of the cyclotron-Bloch dynamics for $F>F_{cr}$ is the strong dependence of the rate of spreading on the field direction or, what is the same, on the ratio between the two Bloch frequencies. The 1D system (\ref{5}) fairly  reproduces this dependence. As an example, in Fig.~\ref{fig2} we depict the coefficient $A$ for the asymptotic linear grows of the wave-packet dispersion, $\sigma(t)\approx At$, as a function of the frequency $\omega$ for $\omega_x/\omega_y=0$ and  $\omega_x/\omega_y=1$. In the former case, the rate of ballistic spreading is seen to approach the constant value $A=J_x/\sqrt{2}\hbar$, while in the latter case it decreases as 
\begin{equation}
\label{8}
A= \frac{J_x}{2\hbar} \left(\frac{J_y}{\hbar\omega}\right)^\nu  \;,
\end{equation}
where $\nu=1$. In the general case of arbitrary rational ratio $\omega_x/\omega_y=r/q$ the exponent is $\nu=r+q-1$ and one has to evolve the system for algebraically large times to reach the asymptotic regime. Finally, for irrational $\omega_x/\omega_y$ there is no asymptotic linear growth in $\sigma(t)$ but oscillations in time (see Fig.~\ref{fig3}).

\section{Finite evolution times}

\begin{figure}
\center
\includegraphics[width=10cm,clip]{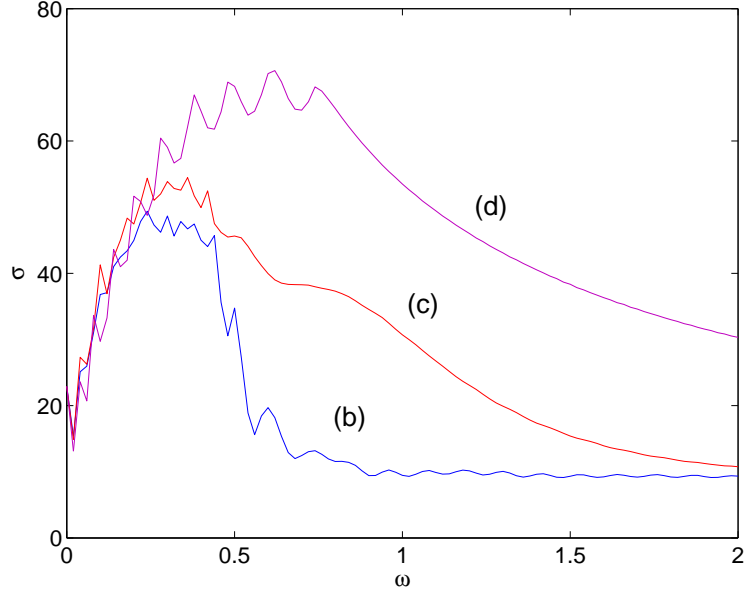}
\caption{The wave-packet dispersion at $t=30 T_J$ as the function of $\omega$ for $\omega_x/\omega_y=1$ (d), $\omega_x/\omega_y=1/3$ (c), and $\omega_x/\omega_y=(\sqrt{5}-1)/4\approx 0.309$ (b).}
\label{fig4}
\end{figure}

\begin{figure}
\center
\includegraphics[width=10cm,clip]{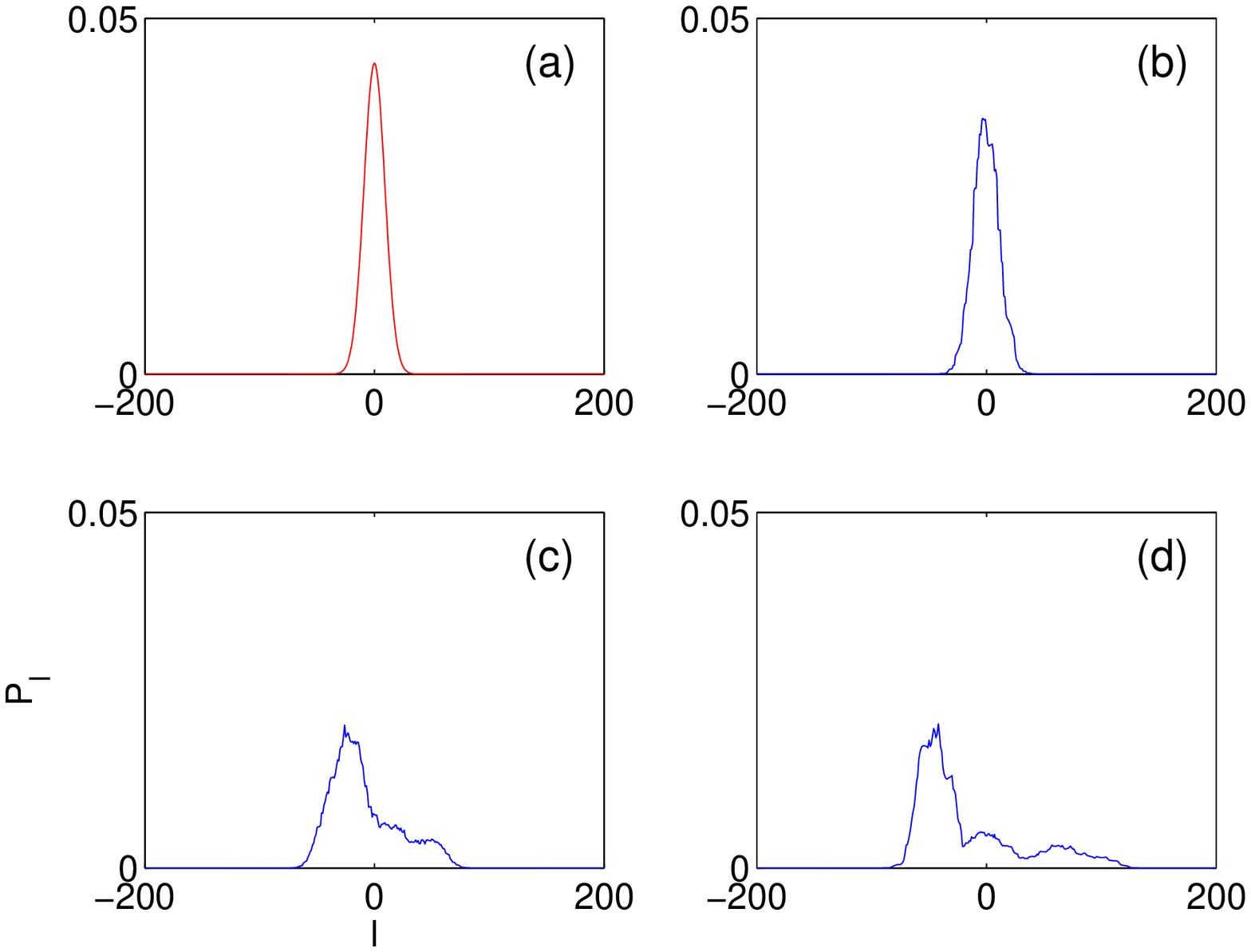}
\caption{Population of the lattice sites at the end of numerical simulation for $\omega=1$. Panels (b-d) correspond to the three cases considered in Fig.~4, the panel (a) shows the initial packet.}
\label{fig5}
\end{figure}

In the preceding sections we discussed asymptotic dynamics of the system. Clearly, in a laboratory experiment the system evolution is restricted to some maximal time interval, which may be not large enough to speak about the asymptotic regime. Nevertheless, all effects mentioned above are well observed also for short evolution times. To support this statement Fig.~\ref{fig4} shows the wave-packet dispersion at  $t=30 T_J$ as the function of $\omega$ for three different ratios between driving frequencies. Two dynamical regimes, which are separated by the critical frequency (\ref{6}), can be easily identified and for $\omega>\omega_{cr}$ one clearly sees the difference between rational and irrational $\omega_x/\omega_y$. In addition to Fig.~\ref{fig4}, the panels (b-d) in Fig.~\ref{fig5} depict populations of the lattice sites at the end of numerical simulations for $\omega=1$. (Tiny wiggling  of the curves is an artifact due to the Monte-Carlo method of simulating the dynamics of an incoherent packet.)  We note that particular shapes of the packets seen in the figure are rather sensitive to variations of the system parameters and the evolution time. This sensitivity, however, disappears for integrated characteristics like the wave-packet dispersion.

\section{Conclusion} 

We have studied numerically the dynamics of non-interacting cold atoms in the driven 1D optical lattice with a particular driving. This driving assumes the presence of a static force, which we characterize by the parameter $\omega_x$ -- the Bloch Frequency, and a shallow secondary lattice with a larger period $d=a/\alpha$, which moves at constant velocity $a\omega_y$ relative to the deep primary lattice. It is shown that this system well reproduces many features of the cyclotron-Bloch dynamics of the quantum particle in a 2D lattice. In particular, we find two qualitatively different dynamical regimes of the 1D system, depending on the driving frequencies. In the first regime (slow driving, $\omega\equiv\sqrt{\omega_x^2+\omega_y^2}<\omega_{cr}$) a cloud of non-condensed atoms spreads ballistically across the lattice with a rate proportional to the frequency $\omega_y$. In the second regime (fast driving, $\omega>\omega_{cr}$) the size of the atomic cloud oscillates in time if $\omega_x/\omega_y$ is an irrational number, while for rational $\omega_x/\omega_y=r/q$ these oscillations are accompanied by slow ballistic spreading with a rate inversely proportional to the frequency $\omega$ in the power $\nu=r+q-1$.

We conclude the paper by a remark concerning the commensurability condition between the lattice periods, i.e., the rationality condition on the parameter $\alpha$.  This condition is known to play a crucial role in the case of stationary lattices \cite{Aubr80,Roat08}. Unlike this situation, for driven lattices we did not find the commensurability condition on the lattice periods to be of any importance, although we do not exclude a possibility that it might be important in some particular situations.

\vspace*{5mm}
\noindent
{\it Acknowledgments}\\
This work was partially supported by Russian Foundation for Basic Research, grant RFBR-10-02-00171-a.

\begin{thebibliography}{10}
\bibitem{Hofs76}
D.~R.~Hofstadter,
Phys. Rev. B {\bf 14}, 2239 (1976).

\bibitem{Peie33}
R.~E.~Peiers, 
Z. Phys. {\bf 80}, 763 (1993).

\bibitem{Naka93}
T.~Nakanishi, T.~Ohtsuki, and M.~Saitoh,
J. of Phys. Soc. Japan {\bf 62}, 2773 (1993).

\bibitem{51} 
M.~Gl\"uck, F.~Keck, A.~R.~Kolovsky, and H.~J.~Korsch,
Phys. Rev. Lett. {\bf 86}, 3116 (2001).

\bibitem{Naka95}
T.~Nakanishi, T.~Ohtsuki, and M.~Saitoh,
J. of Phys. Soc. Japan {\bf 64}, 2092 (1995).


\bibitem{Roat08}
G. Roati, C. D' Errico, L. Fallani, M. Fattori, C. Fort, M. Zaccanti, G. Modugno, M. Modugno, and M. Inguscio,
Nature {\bf 453}, 891 (2008).

\bibitem{Aubr80}
S.~Aubry and G.~Andr\'e,
Ann. Israel. Phys. Soc. {\bf 3}, 133 (1980).

\bibitem{Harp55}
P.~G.~Harper,
Proc. Phys. Soc. A {\bf 68}, 874 (1955).


\bibitem{84} 
A.R.Kolovsky,
Europhys. Lett. {\bf 93}, 20003 (2011).

\bibitem{85}
A.~R.~Kolovsky and G.~Mantica,
Phys. Rev. E {\bf 83}, 041123 (2011).

\bibitem{preprint}
I.~Chesnokov, A.~R.~Kolovsky and G.~Mantica,
in preparation.

\bibitem{remark1}
We note that in the experiment \cite{Roat08} the authors used a mirror to create the standing waves. In this work we assume the other scheme of the experimental setup, where the standing waves are formed by counter-propagating laser beams.   

\bibitem{remark2}
We recall that there is a family of transporting states in the original problem, each representative of which is naturally characterized by its dispersion in the two orthogonal  directions \cite{preprint}.


\end{thebibliography}
\end{document}